\documentclass[%
 reprint,
 amsmath,amssymb,
prb,
floatfix
]{revtex4-1}

\usepackage{graphicx}
\usepackage{dcolumn}
\usepackage{bm}
\usepackage{lipsum} 
\usepackage{amsmath}
\usepackage{braket}
\usepackage{dsfont}
\usepackage{stackengine,graphicx}
\usepackage[percent]{overpic}

\newcommand{\kett}[1]{ | #1 \rangle \rangle}
\newcommand{\braa}[1]{ \langle \langle #1 |}
\newcommand{\braakett}[2]{\langle \langle #1 | #2 \rangle \rangle}
\newcommand{\expec}[1]{\langle #1 \rangle}

\usepackage{color}
\usepackage{ulem}

\begin{document}


\title{Counting statistics of dark-state transport through a carbon nanotube quantum dot}

\author{Nathan Ho}
 
\author{Clive Emary}%

\affiliation{Joint Quantum Centre Durham-Newcastle, School of Mathematics, Statistics and
Physics, Newcastle University, Newcastle upon Tyne NE1 7RU, UK}

\date{\today}

\begin{abstract}
In a recent experiment [A.\,Donarini \textit{et al.},~Nat.\,Comms \textbf{10}, 381 (2019)], electronic transport through a carbon nanotube quantum dot was observed to be suppressed by the formation of a quantum-coherent ``dark state''.  In this paper we consider theoretically the counting statistics and waiting-time distribution  of this dark-state-limited transport. We show that the statistics are characterised by giant super-Poissonian Fano factors and long-tailed waiting-time distributions, both of which are signatures of the bistability and extreme electron bunching caused by the dark state.
\end{abstract}

\maketitle

\section{Introduction}
In quantum optics, coherent population trapping is a phenomenon in which coherent illumination drives an electron into a particular superposition of orbital states --- a dark state --- that is decoupled from the light fields \cite{Alzetta,Arimodo, Whitley}.
In Ref.~\onlinecite{Michaelis}, it was suggested that an all-electronic analogue of this effect should exist in the transport through nano-electronic systems such as a triple quantum dot. In this scenario, it is coherent tunneling between electronic states that permits the formation of a trapped state, and the ``darkness'' of this state is manifested as the suppression of electronic current through the system.
Recently,  Donarini \textit{et al.} \cite{Donarini19} reported the observation of current suppression in the transport through a carbon nanotube quantum dot (CNT-QD) and explained this effect as arising through the presence of a dark state formed by the superposition of longitudinal-orbital-momentum states in the nanotube.

In this paper we report on calculations of the counting statistics \cite{LeviLeso,Levitov1996,Bagrets,Gustav2,Gustav1,Fujisawa,Kurzmann2019} and waiting time distribution \cite{TBrandes} of the nano-tube model introduced in Ref.~\onlinecite{Donarini19}.  The counting statistics is a well-established tool for obtaining information about transport processes beyond that which is available from measurements of the mean current alone \cite{LeviRez,DLoss,Kiesslich}. Here we use the counting statistics formalism to investigate the current noise and skewness in particular.  The waiting time distribution, i.e. the distribution of times between consecutive electron-tunneling events, gives insight that is complementary to that obtained with the counting statistics \cite{Emary12}.

The counting statistics of other transport dark-state models such as the triple quantum dot have previously been reported \cite{Groth06,Emary07,Poeltl09,Weymann2011,Dominguez2010,Dominguez2011,Poeltl13,Zhao2014} with the dark state generally leading to super-Poissonian statistics associated with electron bunching.  For the CNT-QD model here, we also obtain super-Poissonian statistics.  However, the degree of this effect depends very strongly on a parameter $\Delta \phi$ which describes the phase difference between tunneling states of the source and drain leads.  Indeed, for small $\Delta \phi$, we report a diverging noise and a skewness that both diverges and changes sign.
In this regard, we conclude that the behaviour of the model is similar to that of the the Aharonov-Bohm interferometer models discussed in Refs.~\onlinecite{Urban09,Li09}.
We also study in detail the effect on the counting statistics of two  mechanisms that break the coherent population trapping, namely relaxation and a Lamb-shift precession.

Concerning the  waiting time distribution, we show that the presence of the dark state gives rise to distributions with extremely long tails.  Moreover,  under certain conditions, we find that the waiting time distribution shows oscillations when the Lamb-shift is the dominant dark-state unblocking mechanism.   This gives, in principle at least, a means though which the dark-state-breaking mechanism could be identified.

Finally, by using the parameters and voltage-dependence of the Lamb shifts from the CNT-QD experiment of Ref.~\onlinecite{Donarini19}, we outline how the above features would appear in this experiment.  We discuss how noise measurements can be useful in estimating critical model parameters, especially $\Delta \phi$, and provide an additional test of whether dark-state physics is indeed responsible for the current rectification in the experiment of Donarini \textit{et al.}.

\section{Model}

\begin{figure}[tb]
    \includegraphics[width=0.9\linewidth]{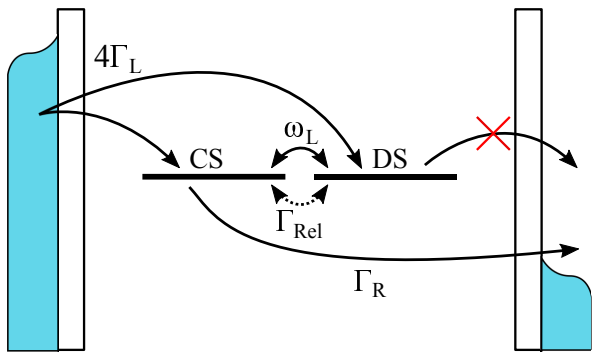}
    \caption{Sketch of the CNT-QD transport model. Electrons enter the system from the left (source) lead with total rate $4\Gamma_L$ into either coupled state (CS) or dark state (DS).  Electrons exit the system into the right (drain) lead at rate $\Gamma_R$ from the CS only. Without anything further, an electron entering the dark state thus becomes trapped, blocking current flow. However, Lamb-shift precession (frequency $\omega_{L}$) and relaxation (rate $\Gamma_\mathrm{Rel}$) transfer electrons between DS and CS, unblocking the system and leading to current flow.
    }
    \label{fig:model_diagrams}
\end{figure}

Our starting point is the model derived by Donarini \textit{et al.} to describe transport within the $N=0,1$ sector of the CNT-QD spectrum \cite{Donarini19}.  This sector consists of the many-body ground state $\ket{0}$ and two degenerate states $\ket{\pm l}$ each with a single excess electron of longitudinal orbital momentum $\pm l$. The spin of the electron only serves to provide degeneracy factors and is otherwise neglected here.  The Coulomb blockade \cite{Averin} prevents the system from being occupied by more than one excess electron in the bias window considered. 

In a high-bias regime, electrons tunnel into the CNT-QD from the left and out to right, see  Fig.~\ref{fig:model_diagrams}.  The respective tunneling rates are $4\Gamma_L$ and $\Gamma_R$, with the factor 4 coming from degeneracy. Due to off-diagonal elements in the tunnel coupling, electrons do not tunnel directly into or out of states $\ket{\pm l}$, but rather into and out of superpositions of them.  Crucially, the relevant basis for tunneling \textit{out of} the dot consists of the states
\begin{equation}
\begin{split}
    \ket{\mathrm{CS}} &\equiv \frac{1}{\sqrt{2}}(e^{i\phi_R}\ket{l} + e^{-i\phi_R}\ket{-l}); \\
    \ket{\mathrm DS} &\equiv \frac{1}{\sqrt{2}}(e^{i\phi_R}\ket{l} - e^{-i\phi_R}\ket{-l}),
\end{split}
\label{eq:CNTstateform}
\end{equation}
where phase $\phi_R$ is a parameter characterising the coupling to the right (drain) lead.  Here, CS denotes the ``coupled state'' and electrons in this state can leave the CNT-QD to the drain.  Conversely,  DS stands for ``dark state'', and this state is decoupled from the drain lead such that electrons entering it can not tunnel out. There exists a similar basis for tunneling \textit{into} the CNT-QD from the left lead.  This is of the same form but with parameter $\phi_L$ instead of $\phi_R$.

The current blocking by the dark state is then driven by the overlap of these two sets of states, and this is governed by the phase difference $\Delta \phi = \phi_L-\phi_R$. When $\Delta \phi=0$ the bases for tunneling through left and right leads are the same. Thus, electrons tunnel from the left into the coupled state $\ket{CS }$, and then tunnel directly out to the right. There is then no dark-state effect in the current flow.
However, when $\Delta \phi \ne 0$, the left and right tunneling bases are different and electrons from the left tunnel into both coupled and dark states.  Once they enter the dark state, electrons can not tunnel out of it and thus remain permanently trapped and current is blocked.

Aside from tunneling, two further mechanisms are taken into account in the model, both of which serve to unblock the dark state.  The first is the precession of the internal states, arising from Lamb-shifts due to the coupling of the leads.  
In Ref.~\onlinecite{Donarini19}, the frequencies $\omega_L$ and $\omega_R$ of these shifts were found to be functions of applied voltages. 
In sections III-V here, we take them as freely-adjustable parameters to eplore the model, and set $\omega_R=\omega_L$ for convenience. In section VI, consider the voltage-dependence of these quantities.
The second unblocking mechanism is relaxation  caused by inelastic processes such as phonon emission/absorption. This drives the internal state of the CNT-QD into the completely mixed state with a rate $\Gamma_\mathrm{Rel}$. 

In the weak-coupling regime, the transport properties of this system can be determined from a quantum master equation of the form 
\begin{eqnarray}
  \dot{\rho} = \mathcal{W}\rho
  \label{EQ:QME}
  ,
\end{eqnarray} 
where $\rho$ is the reduced density matrix of the CNT-QD, and $\mathcal{W}$ is the Liouville super-operator describing all relevant dynamical processes. 
Appendix \ref{AP:W} shows the Liouvillian for the problem at hand in matrix form.
In Appendix \ref{AP:FCS} we outline the counting-statistics formalism for calculating the cumulants of the current $\langle I^k \rangle_c$ (for $k=1,2,3$ here) as well as the  waiting time distribution $w(\tau)$ from master equation~(\ref{EQ:QME}). 

\section{Current}

\begin{figure}[tb]
    \centering
    \begin{overpic}[width = \linewidth, trim={0 1cm 0 0},clip]{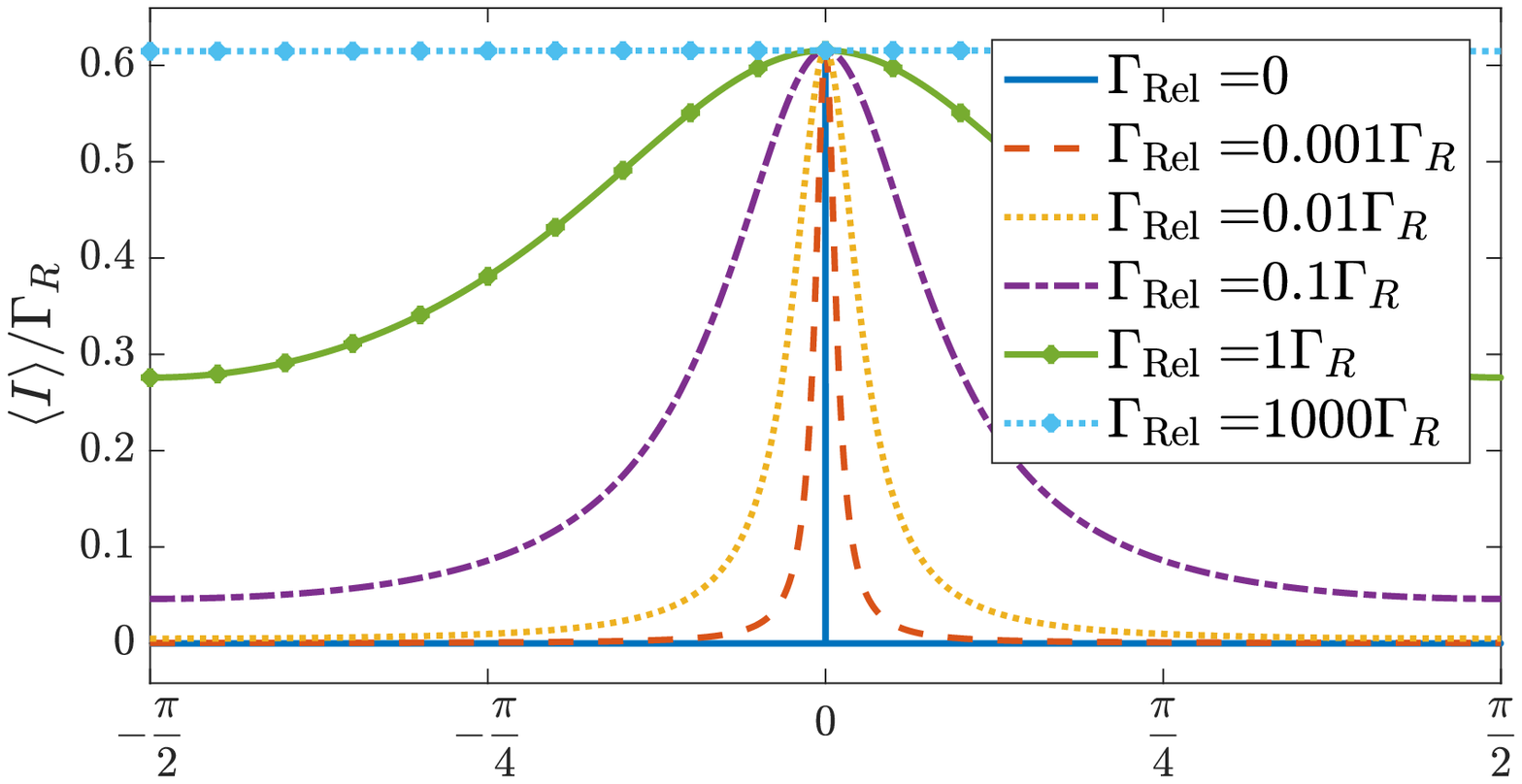}
    \put (11,36) {\large \textbf{(a)}}
    \end{overpic}
    \begin{overpic}[width = \linewidth, trim={0 0 0 0cm},clip]{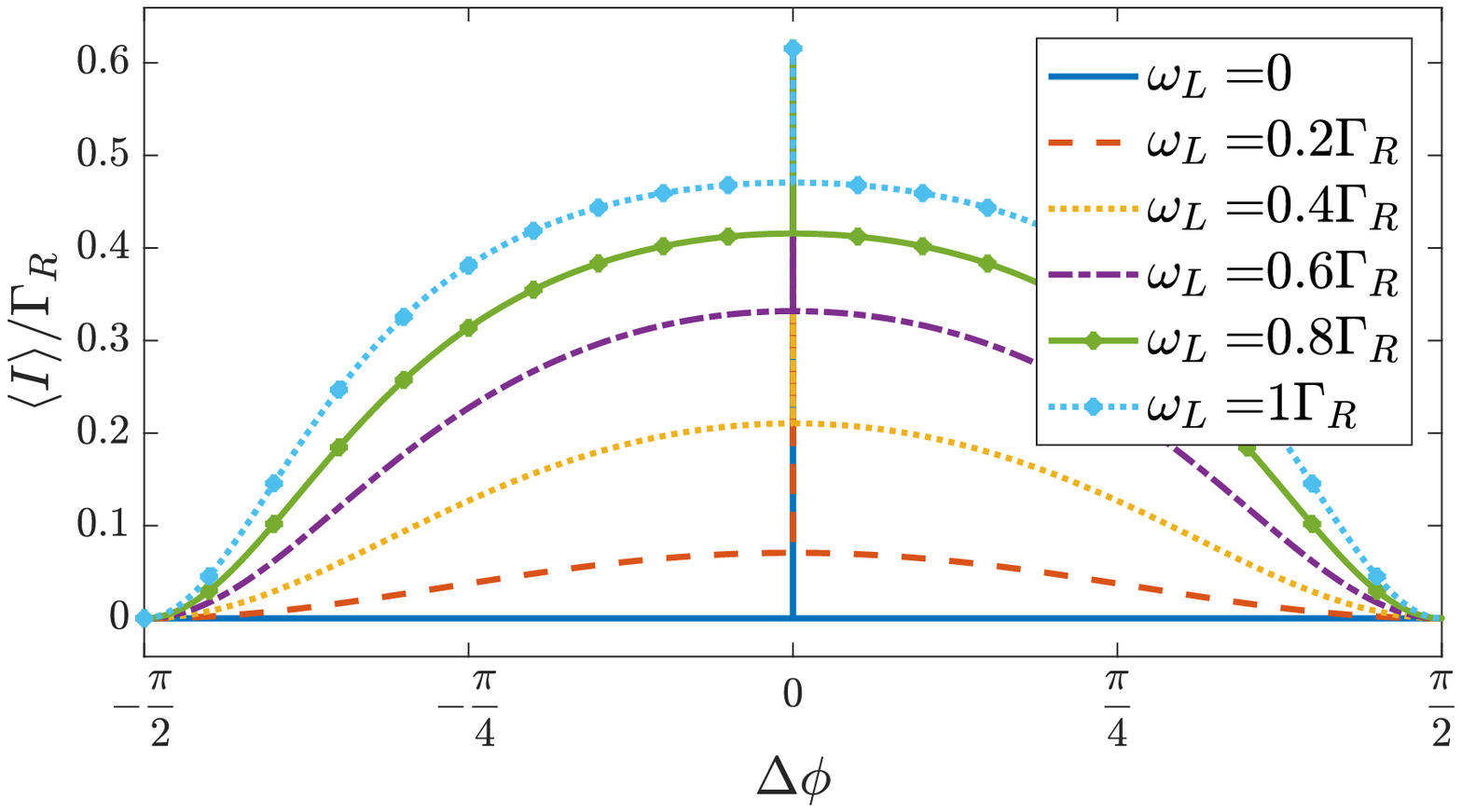}
    \put (11,46) {\large \textbf{(b)}}
    \end{overpic}
    \caption{Mean current $\langle I \rangle$ of the CNT-QD as a function of the tunnel-basis phase difference $\Delta \phi$ with $\Gamma_L = 0.4\Gamma_R$.  Part \textbf{(a)} shows the effect of increasing relaxation rate $\Gamma_\mathrm{Rel}$ with $\omega_L = 0$. Part \textbf{(b)} shows the effect of increasing precession frequency $\omega_L$ with $\Gamma_\mathrm{Rel} = 0$. 
    With $\omega_L = \Gamma_\mathrm{Rel} = 0$ the current is exactly zero for all $\Delta \phi \ne 0$ as the DS blockade is complete. Increasing the strength of either of the unblocking interactions increases the current with largest values for $\Delta \phi\xrightarrow{}0$.
    }
    \label{fig:IPhi}
\end{figure}

The impact of the DS on transport through the CNT-QD is immediately seen in the mean current, $\langle I \rangle$. The results we obtain are consistent with those of  Ref.~\onlinecite{Donarini19}, but here we separate out the influence of the two unblocking mechanisms.

Figure~\ref{fig:IPhi} shows the mean current as a function of the phase difference between tunneling states of the left and right lead, $\Delta \phi \equiv \phi_L - \phi_R$. For any non-zero phase difference, electrons have a finite probability of tunneling into the DS.  In the absence of unblocking mechanisms, this state is decoupled from the right lead, and an electron entering it becomes permanently trapped. This in turn results in complete current suppression, since no further electrons may tunnel into the system due to the Coulomb blockade.

This entrapment is lifted, however, by the relaxation and precession mechanisms. A finite relaxation rate allows for electrons to move from the DS to the CS and then escape into the right lead. Similarly, the Lamb-shift precession causes electrons to oscillate between the CS and DS at a frequency of $\omega_L$ and this allows electrons to escape.  As seen in Fig.~\ref{fig:IPhi}, increasing the strength of either of these processes results in less suppression. The efficacy of precession mechanism in unblocking the system is dependent on the phase difference.

Two special points are evident from these graphs.  The first is $\Delta \phi =0$, where, in the absence of relaxation, the DS is completely decoupled from both left \textit{and} right leads.  In this case, the part of system involved in transport is essentially a single-level system.  The counting statistics of this model are then fully known, as recounted in Appendix \ref{AP:SRL}.  The second special point occurs at $\Delta \phi = \pi/2$, where electrons from the left lead tunnel directly into the DS. With unblocking mechanisms present,  the maximum current always occurs at $\Delta \phi =0$, and the minimum always occurs at $\Delta \phi = \pi/2$. Furthermore, at these points, the current is unchanged by the precession frequency.

\section{Shotnoise and skewness \label{SEC:shotskew}}

The second and third current cumulants are the shotnoise and skewness, respectively.  It is often more useful to discuss the cumulants in terms of their Fano factors, defined as the ratio of the $k^{th}$ cumulant to the first: $F_k = \expec{I^k}_c/\expec{I}$.  The second Fano factor is often just referred to as \textit{the} Fano factor.

\begin{figure}[tb]
    \centering
    \begin{overpic}[width = \linewidth, trim={0 0.83cm 0 0},clip]{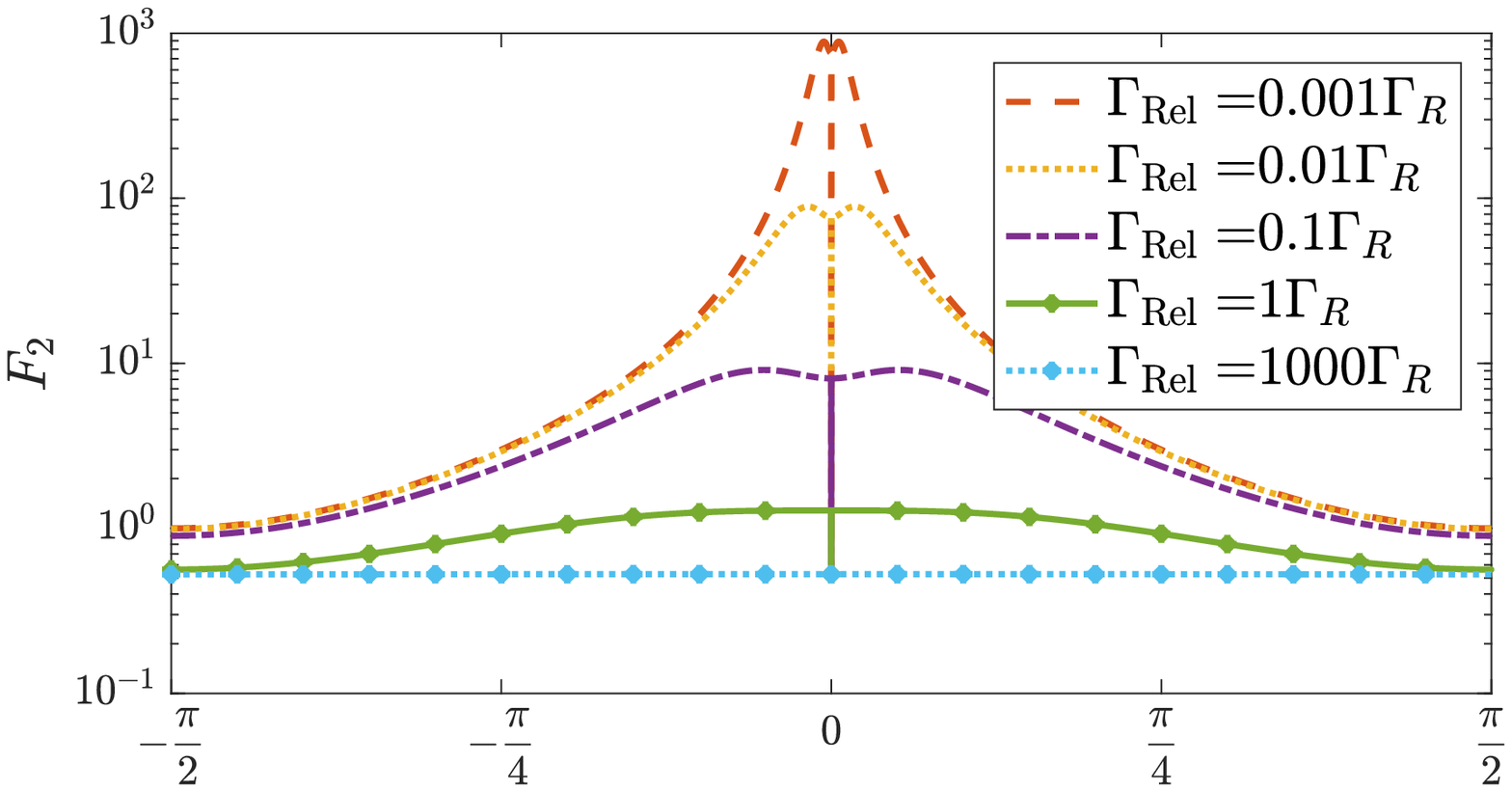}
    \put (12,40) {\large \textbf{(a)}}
    \end{overpic}
    \begin{overpic}[width = \linewidth, trim={0 0 0 0cm},clip]{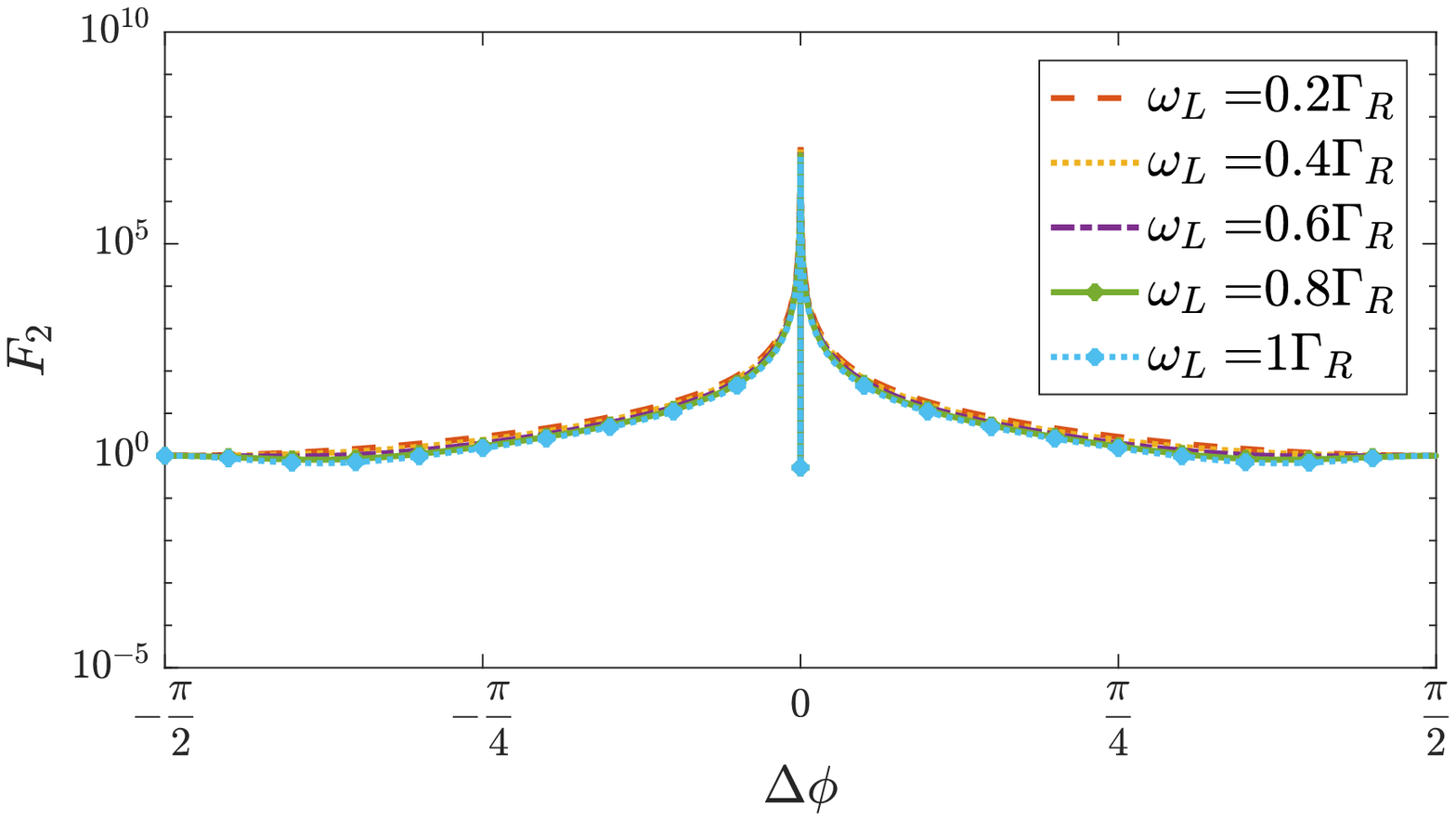}
    \put (12,48) {\large \textbf{(b)}}
    \end{overpic}
    \caption{
    Shotnoise Fano factor $F_2$ as a function of the phase difference $\Delta \phi$ for \textbf{(a)}: increasing relaxation rate $\Gamma_\mathrm{Rel}$ with $\omega_L = 0$, and  \textbf{(b)}: increasing precession frequency $\omega_L$ with $\Gamma_\mathrm{Rel}=0$.
    The most striking feature here is the giant super-Poissonian values assumed by the Fano factor as $\Delta \phi \xrightarrow{}0$. 
    Tunnel rates set as $\Gamma_L = 0.4\Gamma_R$.
    }
    \label{fig:F2vsDelphi_wL}
\end{figure}

Figure \ref{fig:F2vsDelphi_wL} shows the (shotnoise) Fano factor $F_2$ 
as a function of the phase difference for a range of relaxation rates and precession frequencies.  Fig.~\ref{fig:F2vsDelphi_wL}a shows that $F_2$ increases as relaxation decreases for all values of phase difference.  The most striking thing about this plot is that, provided $\Gamma_\mathrm{Rel}\lesssim \Gamma_R$ and the phase different is not near $\pm \pi/2$, the Fano factor assumes a value way in excess of the Poisson value $F_2=1$. And, indeed, as both $\Delta \phi \xrightarrow{}0$ and $\Gamma_\mathrm{Rel}\xrightarrow{}0$, the Fano factor is observed to diverge.

\begin{figure}[tb]
    \centering
    \includegraphics[width = \linewidth]{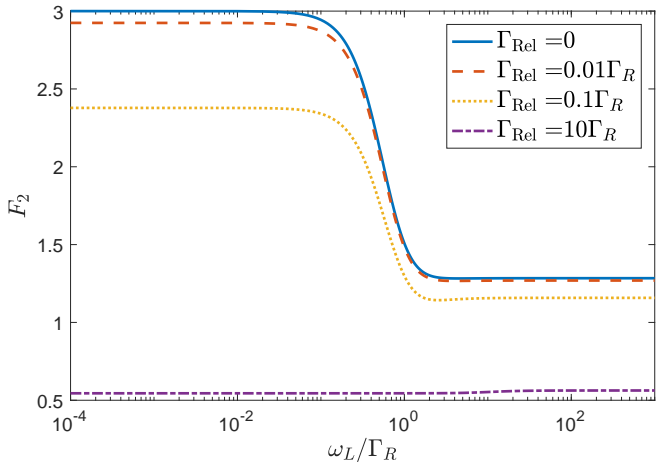}
    \caption{
    The shotnoise Fano factor of the CNT-QD as a function of the precession frequency with varying relaxation rate $\Gamma_\mathrm{Rel}$. Parameters were $\Gamma_L = 0.4\Gamma_R $ and $\Delta\phi = \pi/4$. At low relaxation, the Fano factor shows a transition from low to high values as the precession frequency is decreased.
    For both low precession frequency and relaxation rate, the Fano factor tends to a value of $F_2=3$ here. 
    With a high relaxation rate, the system is sub-Poissonian for all precession frequencies.}
    \label{fig:F2wL}
\end{figure}

Figure~\ref{fig:F2vsDelphi_wL}b shows the effect on $F_2$ of changing the precession frequency. Once again, for finite $\omega_L$, giant super-Poissonian values of $F_2$ are observed with divergence occurring for $\Delta \phi \xrightarrow{}0$. For values of $\Delta \phi$ away from the origin, the change in $F_2$ is less drastic than in the case with changing $\Gamma_\mathrm{Rel}$. In Fig.~\ref{fig:F2wL} we plot $F_2$ as a function of precession frequency for several relaxation rates with a fixed phase difference of $\Delta \phi = \pi/4$, far away from the diverging limit. This figure shows that, provided $\Gamma_\mathrm{Rel}$ is small enough, as the precession frequency is decreased, the Fano factor undergoes a transition from a value close to 1 to a value significantly in excess of it (here $F_2 \to 3$ as $\omega_L \to 0$ for $\Gamma_\mathrm{Rel}=0$).  This transition to the higher $F_2$ value is an indicative that the blocking of the DS is starting to play a significant role.
For large $\Gamma_\mathrm{Rel}$ the statistics are sub-Poissonian irrespective of $\omega_L$.  This is as expected because under such circumstances the internal quantum structure of the system becomes irrelevant and the system essentially becomes a single (degenerate) level system.

\begin{figure}[tb]
    \centering
    \begin{overpic}[width = \linewidth, trim={0 0.93cm 0 0},clip]{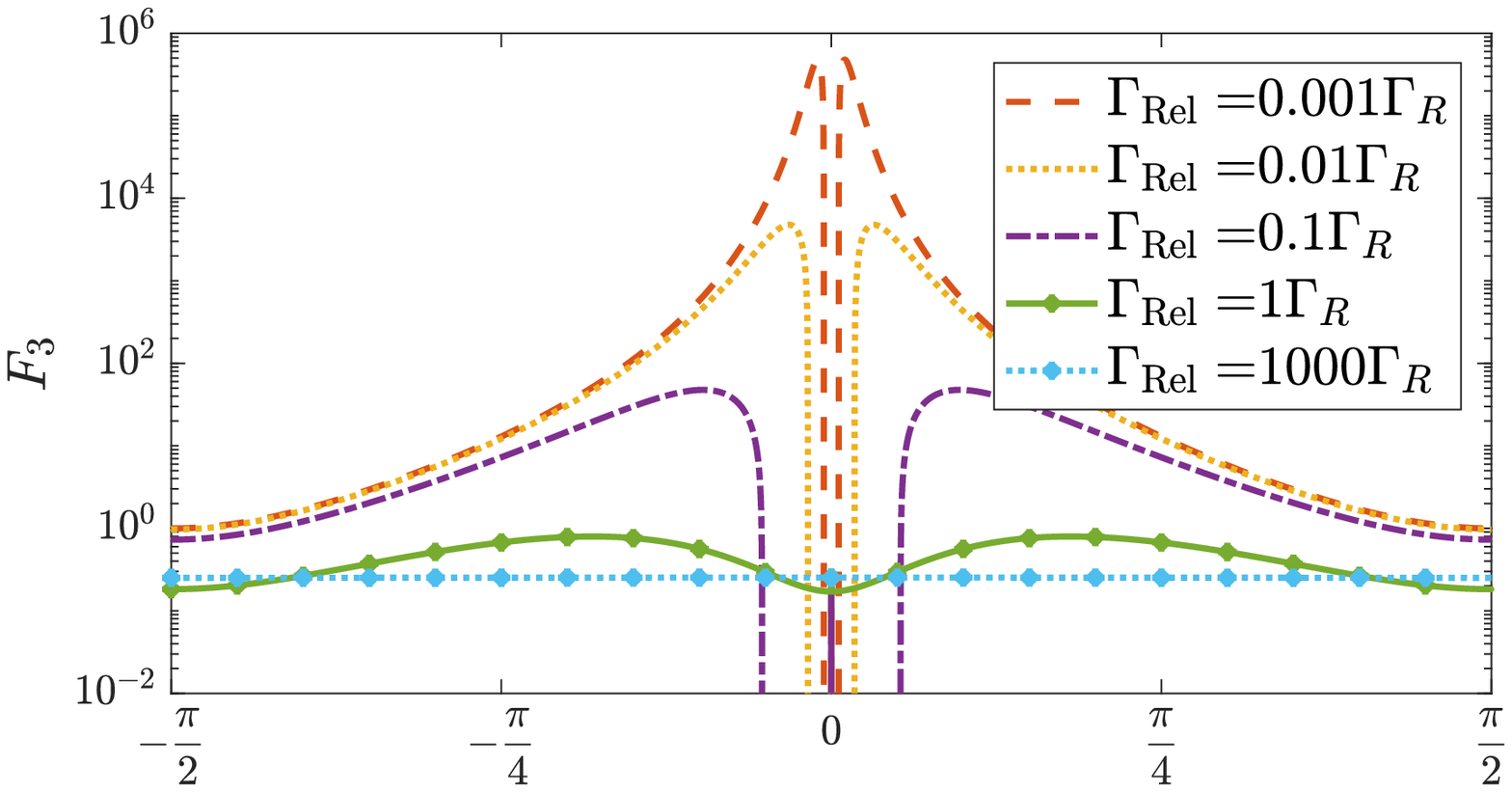}
    \put (12,40) {\large \textbf{(a)}}
    \end{overpic}
    \begin{overpic}[width = \linewidth, trim={0 0 0 0cm},clip]{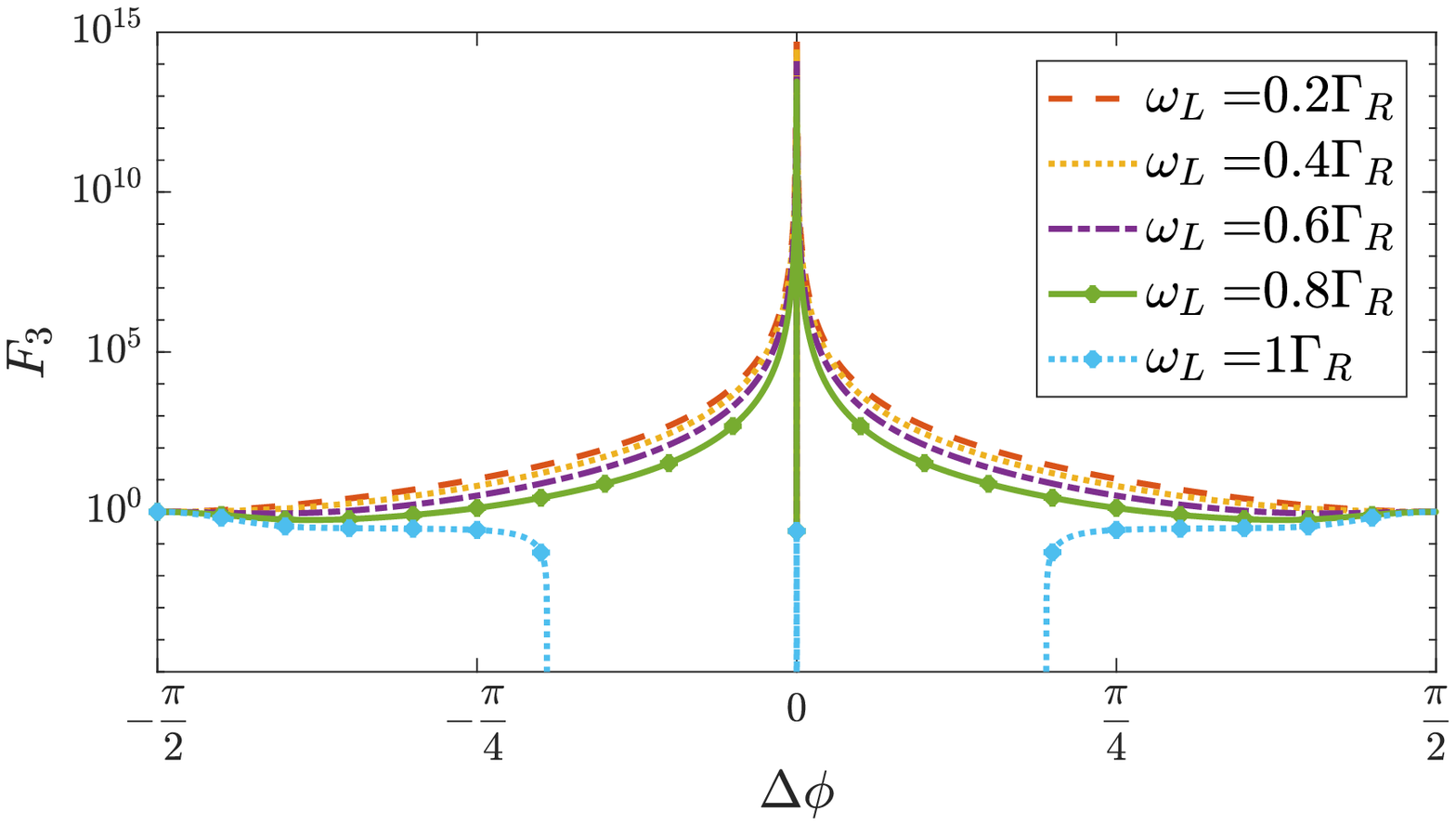}
    \put (12,49) {\large \textbf{(b)}}
    \end{overpic}
    \caption{As Fig.~\ref{fig:F2vsDelphi_wL} but here we plot the skewness Fano factor $F_3$.  As is the case for $F_2$, the third Fano factor is massively super-Poissonian. In addition, for $\Delta \phi \xrightarrow{}0$, $F_3$ becomes negative (which translates as the curves leaving the bottom of these logarithmic plots).
    }
    \label{fig:F3vsDelphi_wL}
\end{figure}

The third (skewness) Fano factor is plotted in Fig.~\ref{fig:F3vsDelphi_wL} as a function of $\Delta \phi$.  Once again we observe giant super-Poissonian values, even larger than those seen with $F_2$ for the same parameters.  As $\Delta \phi \xrightarrow{} 0$, $F_3$ becomes negative and for $\Gamma_\mathrm{Rel}\to 0$ or $\omega_L \to 0$ diverges as $F_3 \to -\infty$.

\section{Effective models \label{SEC:effmodels}}

We now discuss two effective models that allow us to explain many of the features of the foregoing results, as well as to connect with previous studies in the literature.  We concentrate on the $\omega_L=0$ case with relaxation the dominant unblocking mechanism.

Instead of using the quantum master equation, Eq.~(\ref{EQ:QME}), we can alternatively describe the system with a rate equation involving the populations of the three states $\ket{0}$, $\ket{CS}$ and $\ket{DS}$. In this picture, electrons tunnel into the CS with a rate $4\Gamma_L \cos^2(\Delta \phi)$ and into the DS with a rate $4\Gamma_L \sin^2(\Delta \phi)$. Without relaxation, tunneling to the right lead only occurs from the CS with rate $\Gamma_R$. However, when $\Gamma_\mathrm{Rel}$ is finite, electrons can leave the system by first relaxing into the coupled state and then tunneling out. When $\Gamma_\mathrm{Rel} \ll \Gamma_R$ the speed of this process is limited by the relaxation step and we can write this unblocking step as an effective out-tunneling from the DS to the state $\ket{0}$ at a rate $\Gamma_\mathrm{eff} \approx \Gamma_\mathrm{Rel}$.
Thus, we describe the system with $\chi$-resolved rate equation $\dot{P} = \mathcal{W}(\chi)P$ where
\begin{equation*}
    \mathcal{W}(\chi) = 
    \begin{pmatrix}
    -4\Gamma_L  & \Gamma_Re^{i\chi} & \Gamma_\mathrm{eff}e^{i\chi} \\
    4\Gamma_L \cos^2 (\Delta \phi) & -\Gamma_R & 0 \\
    4\Gamma_L \sin^2 (\Delta \phi) & 0 & -\Gamma_\mathrm{eff} \\
    \end{pmatrix},
\end{equation*}
and where, in the $\chi\to0$ limit, $P$ is the vector of populations of the $\ket{0}$, $\ket{CS}$ and $\ket{DS}$ states.

With this simplified model, exact expressions for the current cumulants are possible.  Reporting results in the large $\Gamma_R$ limit, we obtain
\begin{equation}
    \langle I \rangle = \frac{4\Gamma_\mathrm{eff}\Gamma_L\csc(\Delta \phi)^2}{4\Gamma_L+\Gamma_\mathrm{eff}\csc(\Delta\phi)}
    ,
\end{equation}
and
\begin{equation}
    F_2 = \frac{\Gamma_\mathrm{eff}^2 + 10\Gamma_L^2 - 2\Gamma_L^2 [4\cos(2\Delta \phi) + \cos(4\Delta \phi) ]}{[\Gamma_\mathrm{eff}+2\Gamma_L-2\Gamma_L\cos(2\Delta \phi)]^2}. 
\end{equation}
The corresponding expression for the skewness is rather cumbersome and not especially illuminating. Good agreement is found between the full numerics and these expressions in the appropriate regime.

This effective model then permits us to make immediate connection with the dynamical channel blockade models of Belzig and co-workers \cite{Cottet2004,Belzig05}. Indeed, with  $\Delta \phi = \pi/4$ as in Fig.~\ref{fig:F2wL}, the rates of tunneling into each of the CS and DS become equal and the model here becomes identical with that of Ref.~\onlinecite{Belzig05}.  This then explains the  Fano factor value of $F_2=3$ in the limit $\omega_L \xrightarrow{}0$ for $\Gamma_\mathrm{Rel}=0$ as arising from electron bunches with multiples of 3 electrons per bunch.  This then also matches with the Fano factor from found by Groth \textit{et al.} \cite{Groth06} for the triple-quantum dot model.
For $\Delta \phi \ne \pi/4$, the tunnel rates into the two states become unequal, and this then significantly changes the bunching properties of the current flow.

Our second effective model provides a simple explanation of the diverging Fano factors, as well as their signs. For $\Gamma_\mathrm{Rel}\approx 0$, our transport system is essentially bistable \cite{Schaller10}: in one of its steady states (the DS) the system does not conduct; in the other, it does and admits a mean current, $\langle I \rangle_0$, say.  The probability distribution for the number of transferred charge will therefore be approximately   
$P(n,t) = (1-p) \delta_{n,0} + p \delta_{n,n_0}$ where $p$ is the probability that we find ourselves in the conducting channel and $n_0 = \langle I \rangle_0 t$ is the mean number of transferred charges in time $t$ if we do. If $t$ is large, we are justified in ignoring the small fluctuations in the charge numbers of zero and $n_0$.  The cumulant generating function for this model reads $\mathcal{F}(\chi,t) = \ln (1-p + p e^{i n_0 \chi})$, which means we have a Bernoulli distribution with ``pay-off'' $n_0$.  The mean current is 
$ \langle I \rangle = 
  t^{-1} \left. \partial\mathcal{F}/(\partial (i\chi)) \right|_0 = p \langle I \rangle_0
$, which is the mean current of the conducting state multiplied the probability of obtaining that state.  Then, the first two Fano factors read $F_2 = \langle I \rangle_0 t (1-p)$ and $F_3= \langle I \rangle_0^2  t^2 (1-p) (1-2p)$.  In the asymptotic limit, $t \to \infty$, the Fano factors diverge as $F_k \sim t^{k-1}$. Moreover, whilst the sign of $F_2$ is manifestly positive (since $1-p>0$), the skewness will be positive for $p<1/2$ and negative for $p>1/2$ due to the factor $(1-2p)$.  Indeed, this simple model suggests that negative skewness is associated with a bistable situation when the probability to find the conducting channel is large, and that the skewness will transfer to being positive as the weight of the blocking channel increases.  This is the behaviour observed in the DS model here.

\section{Waiting-Time Distribution}

\begin{figure}[tb]
  \centering
  \includegraphics[width=\linewidth]{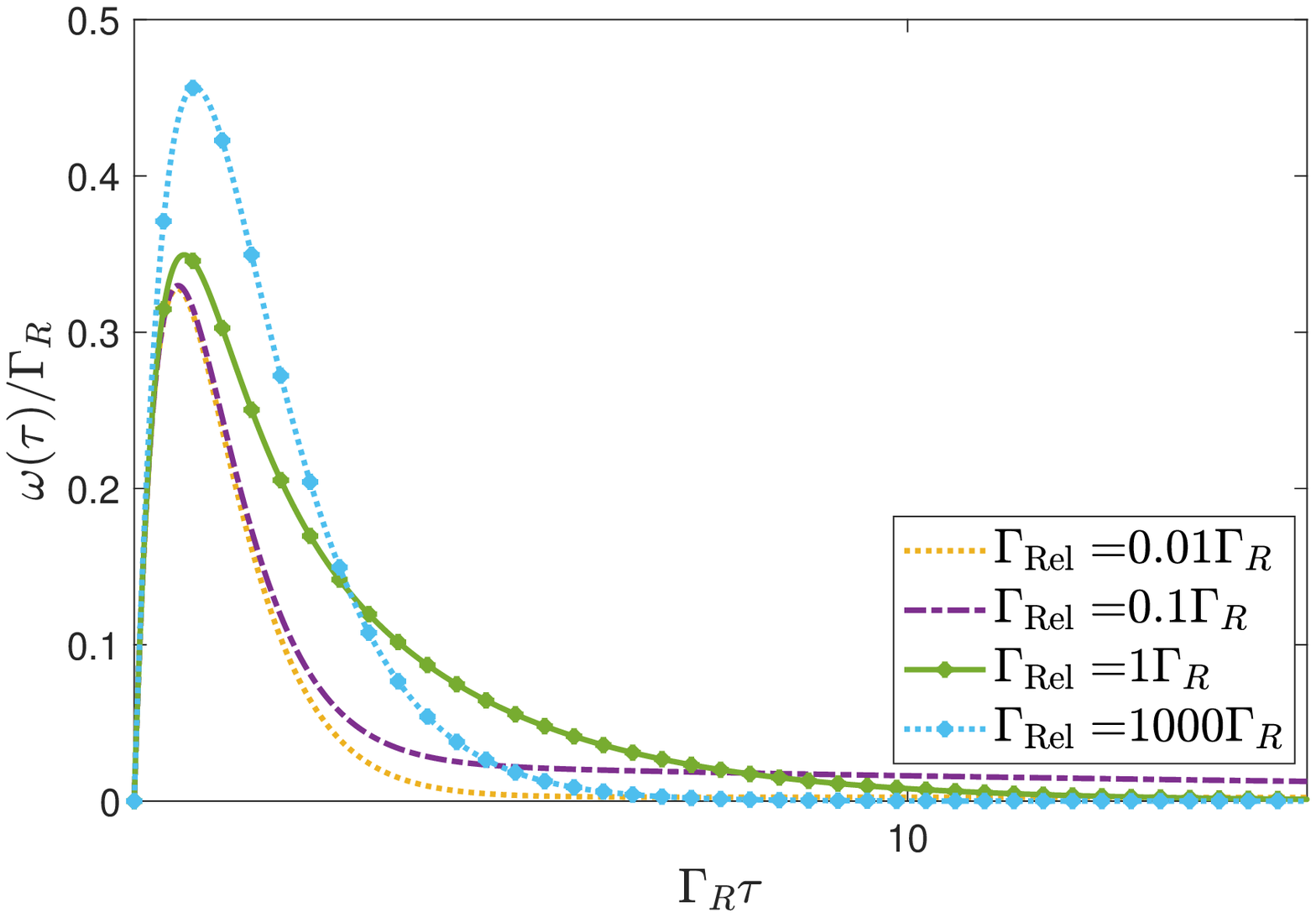}\llap{\makebox[5cm][l]{\raisebox{2.675cm}{\includegraphics[width = 0.55\linewidth,clip=true]{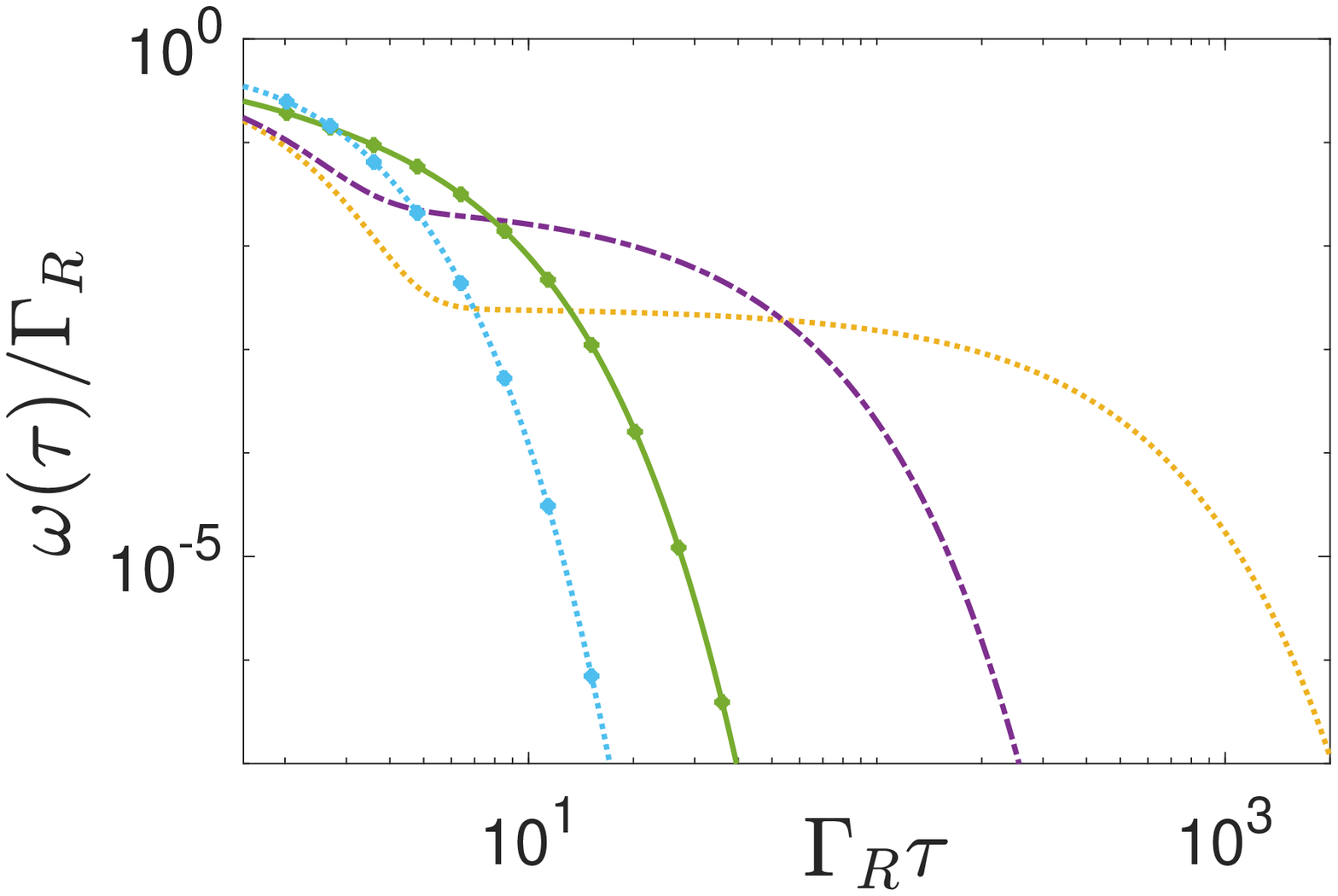}}}} \\
  \caption{Waiting time distributions for various $ \Gamma_\mathrm{Rel}$ with $\omega_L = 0$, $\Gamma_L = 0.4\Gamma_R $ and $\Delta \phi = \pi/4$. Inset shows the same data on a logarithmic scale, which highlights the long tail of the distribution induced by the dark state.}
  \label{fig:wtd_vary_GRel}
\end{figure}

The waiting time distribution, $\omega(\tau)$ gives the probability of waiting a time $\tau$ between consecutive ``jumps'' of an electron out of the system. Fig.~\ref{fig:wtd_vary_GRel} shows the waiting time distribution for the same parameters as Fig.~\ref{fig:IPhi}a with the phase difference fixed at $\Delta \phi = \pi/4$.  Whilst on a linear scale, the distribution $w(\tau)$ looks similar to that which would be obtained from a single-level quantum dot \cite{TBrandes}, on a log scale (Fig.~\ref{fig:wtd_vary_GRel} inset), we see that the distribution possesses an extremely long tail, and the lower the relaxation rate, the longer the tail becomes.   This tail is due to the presence of the DS which results in electron being trapped for long times before exiting the system.  The separation of time scales can be extreme. For $\Gamma_\mathrm{Rel} /\Gamma_R= 0.01$, for example, the main peak of the tunneling dynamics is over after $\tau \approx 5 \Gamma^{-1}$ whereas the bulk of the tail extends out to a time of $\tau \approx 10^3 \Gamma^{-1}$.
We note that the waiting-time distribution shows no particular trace of the change in sign of the third Fano factor.

\begin{figure}[tb]
  \centering
  \includegraphics[width=\linewidth]{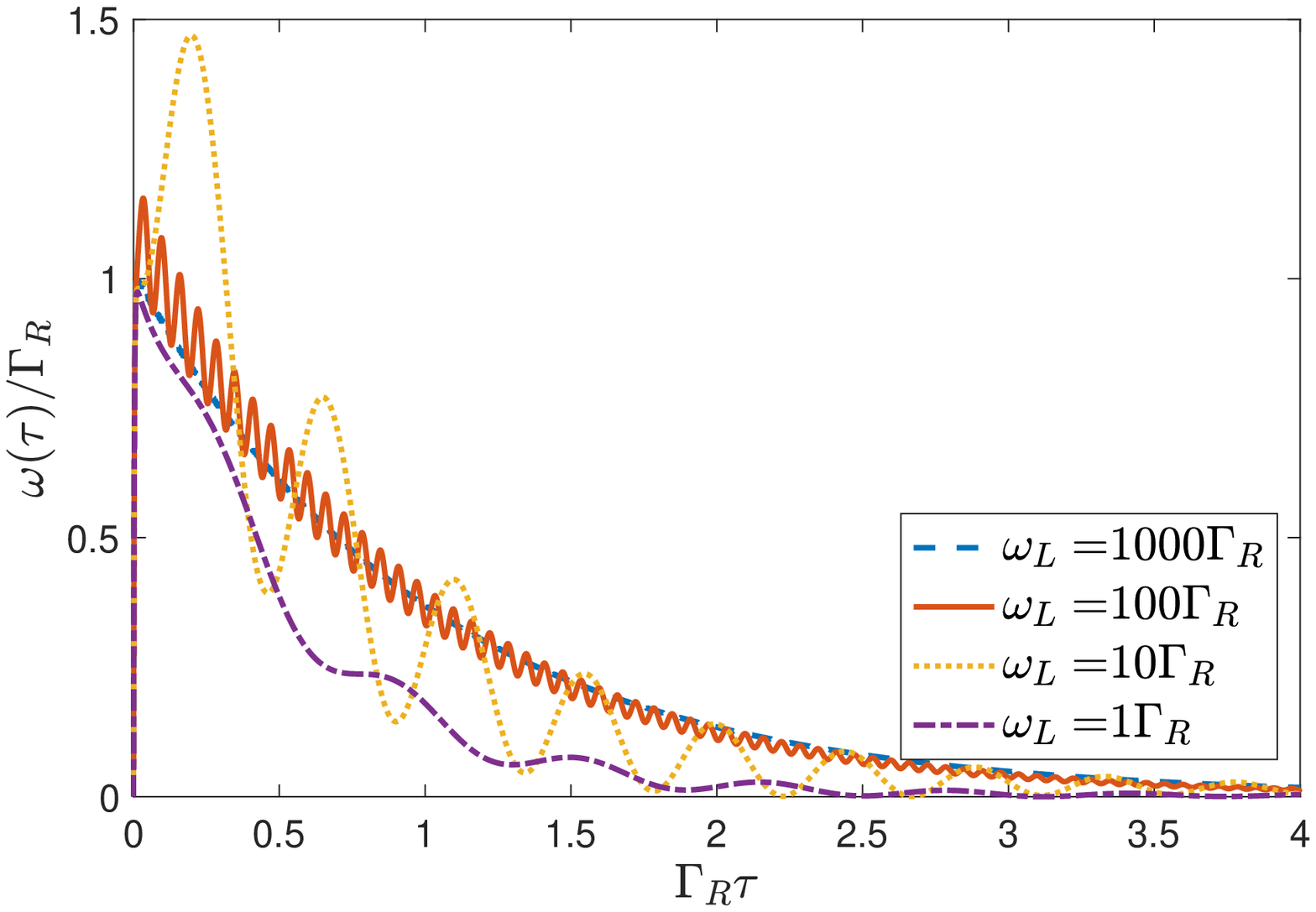}\llap{\makebox[5cm][l]{\raisebox{2.675cm}{\includegraphics[width = 0.55\linewidth,clip=true]{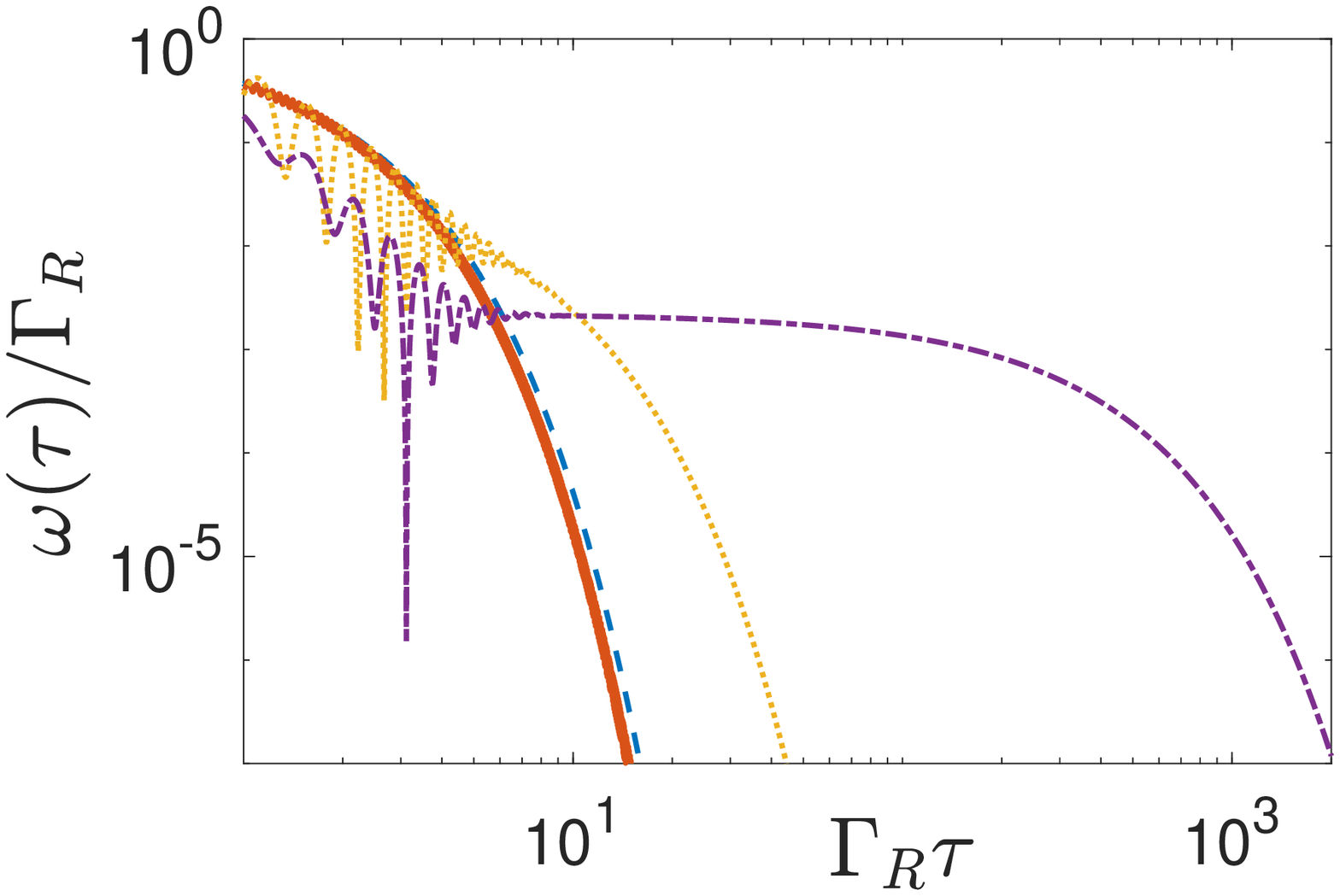}}}} \\
  \caption{
    Waiting time distributions for $\Gamma_\mathrm{Rel}=0$, $ \Gamma_L = 100\Gamma_R$ and $\Delta \phi = \pi/4$ on both linear and logarithmic scales. For these parameters, oscillations due to the Lamb-shift precessions are clearly visible.
  }
  \label{fig:wtd_GRel_0}
\end{figure}

Fig.~\ref{fig:wtd_GRel_0} shows the waiting time distribution when the precesion frequency is finite. In the regime when $\Gamma_L > \Gamma_R$, the precession of the electrons oscillating between the CS and DS imprints oscillations on the waiting time distribution.   As the precession frequency increases, the observed oscillation frequency increases, but the amplitude decreases.  Under these conditions, the waiting time distribution still maintains the extended tail, indicating the continued influence of the DS.

\section{Experimental signatures}
We now consider the appearance of some of the above features in the experimental set-up of Ref.~\onlinecite{Donarini19}.
In the gated CNT-QD set-up, the Lamb shifts are determined by the applied bias $V_B$ and gate $V_G$ voltages via \cite{Donarini19}  ($e=1$)
\begin{equation}
  \omega_\alpha = \frac{\Gamma_\alpha}{\pi}
  \left[
    p_\alpha\left(-V_G\right) - p_\alpha\left ( U - \frac{J}{2} -V_G \right)
  \right]
  ,
\end{equation}
with 
\begin{equation} 
    p_\alpha\left(E\right)  
    = - 
    \mathrm{Re} \,
    \psi
    \left[
      1/2 + i(E-\mu_\alpha)/(2 \pi k_BT)
    \right]
    .
\end{equation}
Here, $\psi$ is the digamma function, $U$ is the QD charging energy, $J$ the exchange interaction strength, and $k_BT$ the thermal energy. The chemical potentials of the leads are set as $\mu_L = -\eta V_B$ on the left and and $\mu_R = (\eta-1)V_B$ on the right, with $\eta$ a parameter to account for an asymmetric bias drop at the two leads.

The model we have hitherto considered is valid for large reverse bias.  Specifically, when $|V_B|$ is larger than all other relevant energy scales, i.e. $k_BT$, $\hbar \omega_\alpha$, etc, and also
$\eta V_B <V_G < (\eta-1)V_B$ such that transitions lie within the bias window.  The corresponding model for forward bias can be obtained by swapping all ``left'' and ``right'' quantities in Liouvillian  (\ref{eq:CNTLiouvillian}). This forward bias calculation is then valid at high bias with $(\eta-1) V_B <V_G < \eta V_B$.

\begin{figure}[tb]
    \hfill \begin{overpic}[width = \linewidth, trim={0 1.2cm 0 0},clip]{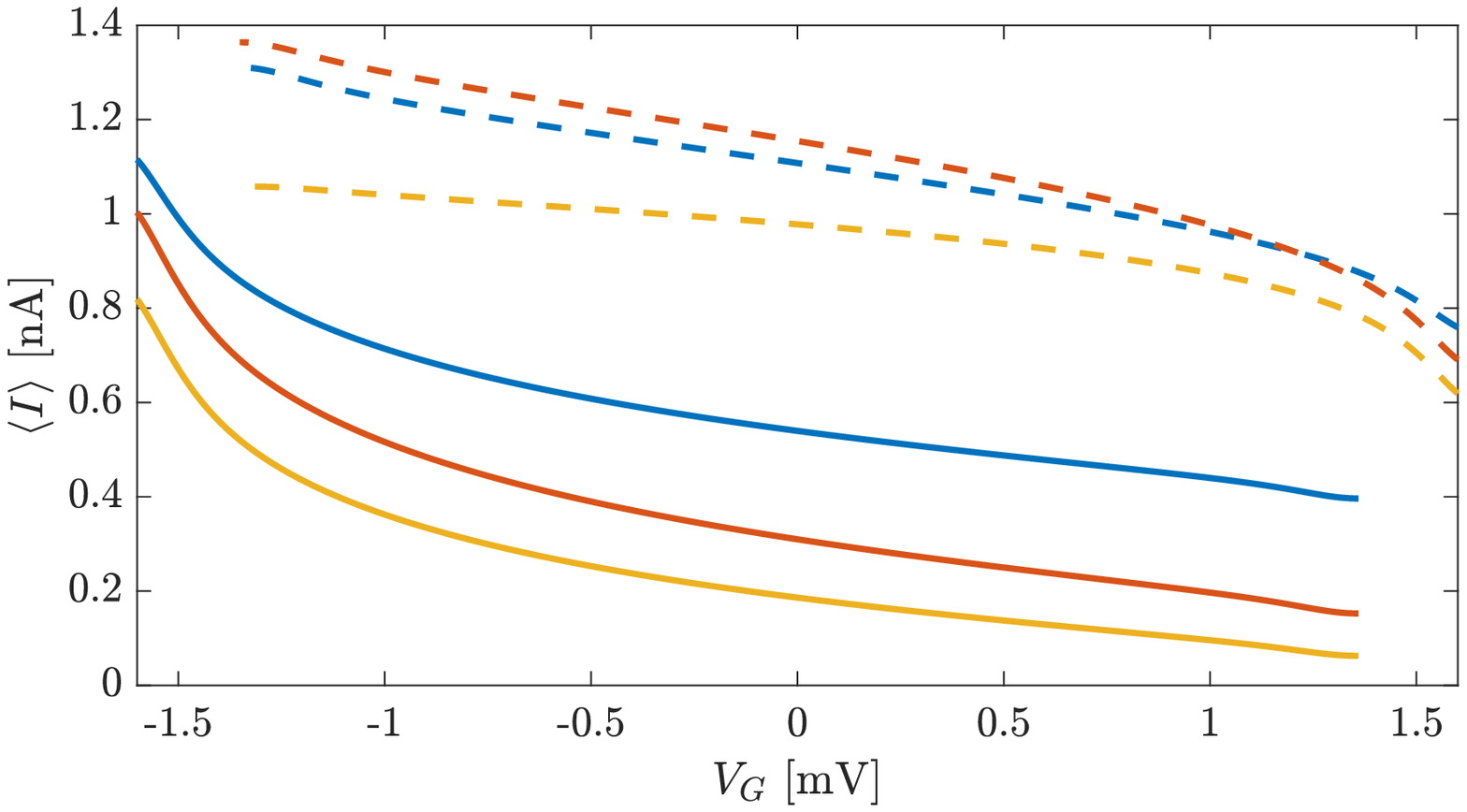}
    \put (12,40) {\large \textbf{(a)}}
    \end{overpic}
    \\
    \hfill\begin{overpic}[width = 0.99\linewidth, trim={0 0 0 0},clip]{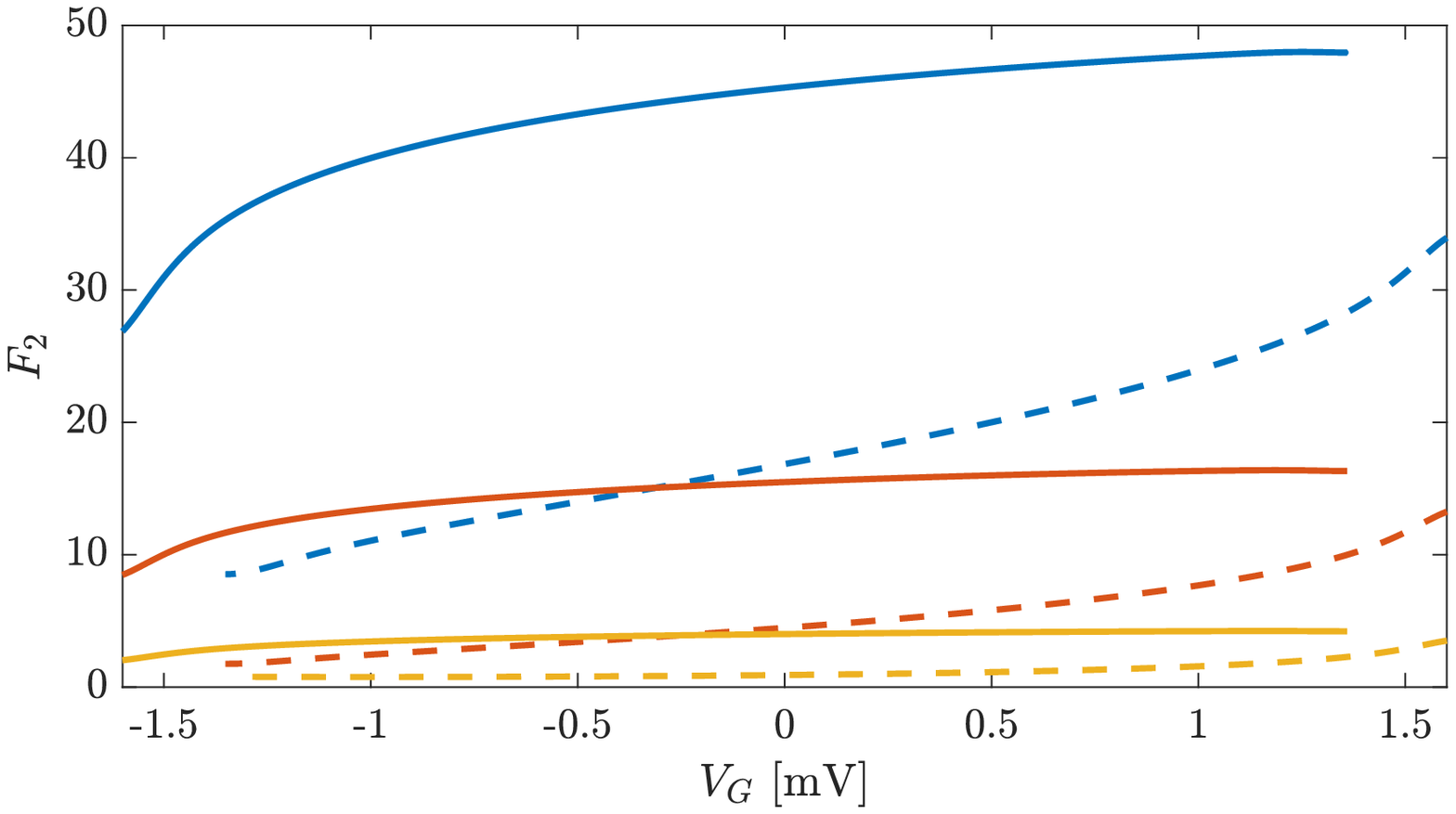}
    \put (11,50) {\large \textbf{(b)}}
    \end{overpic}
  \caption{ 
    \textbf{(a)} Mean current and \textbf{(b)} shotnoise Fano factor $F_2$ for the CNT-QD experiment as a function of gate voltage $V_G$.  Results are shown for three values of the phase difference $\Delta \phi / \Delta\phi_\mathrm{exp} = 0.5,1,2$ (blue, red, yellow curves respectively) where $\Delta \phi_\mathrm{exp} = 0.11 \pi$ is the value found in experiment.  Solid lines show reverse-bias results, dashed lines those for forward bias.
    Strong asymmetry in both current and noise is observed due to the difference in efficacy of the trapping mechanism in the two bias directions.  The Fano factor is seen to be sensitive to changes in the phase difference $\Delta \phi$.
    The bias voltage was set as $V_B = \pm 3$\,mV with other parameters taken from Ref.~\cite{Donarini19}:
    $U = 20$\,meV, 
    $J = 10\,\mu$eV,
    $k_B T = 50\,\mu$eV,
    $\Gamma_L = 4\mu$eV,
    $\Gamma_R = 10\mu$eV,
    $\Gamma_\mathrm{rel} = 0.1\mu$eV, and
    $\eta =0.55$.
  }
  \label{fig:param_dp}
\end{figure}

Figure~\ref{fig:param_dp}a shows the mean current through the CNT-QD as a function of gate voltage for both forward and reverse bias configurations.  We plot results for a typical value of $V_B = \pm 3$\,mV, and for three different phase differences: $\Delta \phi = (0.5,1,2)\times \Delta\phi_\mathrm{exp}$, where $\Delta \phi_\mathrm{exp} = 0.11 \pi$ is the value extracted from experiment \cite{Donarini19}.  
Our results for $\Delta \phi =  \Delta\phi_\mathrm{exp}$ agree with those of Donarini \textit{et al.} and show the rectification that arises from the combination of dark-state trapping and asymmetry in the coupling of the CNT to the leads.

Figure~\ref{fig:param_dp}b shows the predicted shotnoise Fano Factor for the same parameters. We again see a clear distinction between forward- and reverse-bias results, with the reverse-bias Fano factor significantly in excess of that with forward bias.  For the experimentally-determined phase difference, $\Delta \phi =  \Delta\phi_\mathrm{exp}$, the Fano factor at  $V_G=0$ is strongly super-Poissonian, $F_2 = 15.5 $, showing the effect of the dark state.  In forward bias, the $V_G=0$ value is also super-Poissonian, $F_2=5.0$, albeit reduced relative to the reverse bias value.  This difference reinforces that the dark-state trapping is more active in reverse than forward bias.

More significant than the exact value of $F_2$ is its sensitivity to changes in the phase difference $\Delta \phi$.  Ref.~\onlinecite{Donarini19} states that experimentally most of the model parameters can be extracted from the Coulomb diamond properties, and of those that can not, it is the phase difference $\Delta \phi$ and the ratio $\Gamma_L/\Gamma_R$ (for fixed total rate $\Gamma_L+\Gamma_R$) that have the greatest effect in determining the current.
Figure~\ref{fig:param_dp}a shows, however, that changing $\Delta \phi$ by a factor of 2 either way, only leads to relatively small changes in the mean current.
In contrast, the Fano factor changes dramatically with a change in  $\Delta \phi$, as can be seen from Fig.~\ref{fig:param_dp}b. At $V_G=0$ the  reverse-bias Fano factor drops to $F_2 = 4.0$ at $\Delta \phi =  2 \Delta\phi_\mathrm{exp}$ but increases to $F_2 = 45.3$ at $\Delta \phi =  1/2 \Delta\phi_\mathrm{exp}$.  This dramatic change is due to the strong non-linearity of the diverging Fano factor as described in Sec.~\ref{SEC:shotskew}.

This sensitivity of $F_2$, compared with that of the current, should mean that the addition of noise measurements will enable $\Delta \phi$ to be estimated more robustly from experiment than would be the case from current measurements alone.

\begin{figure}[tb]
    \hfill \begin{overpic}[width = \linewidth, trim={0 1.2cm 0 0},clip]{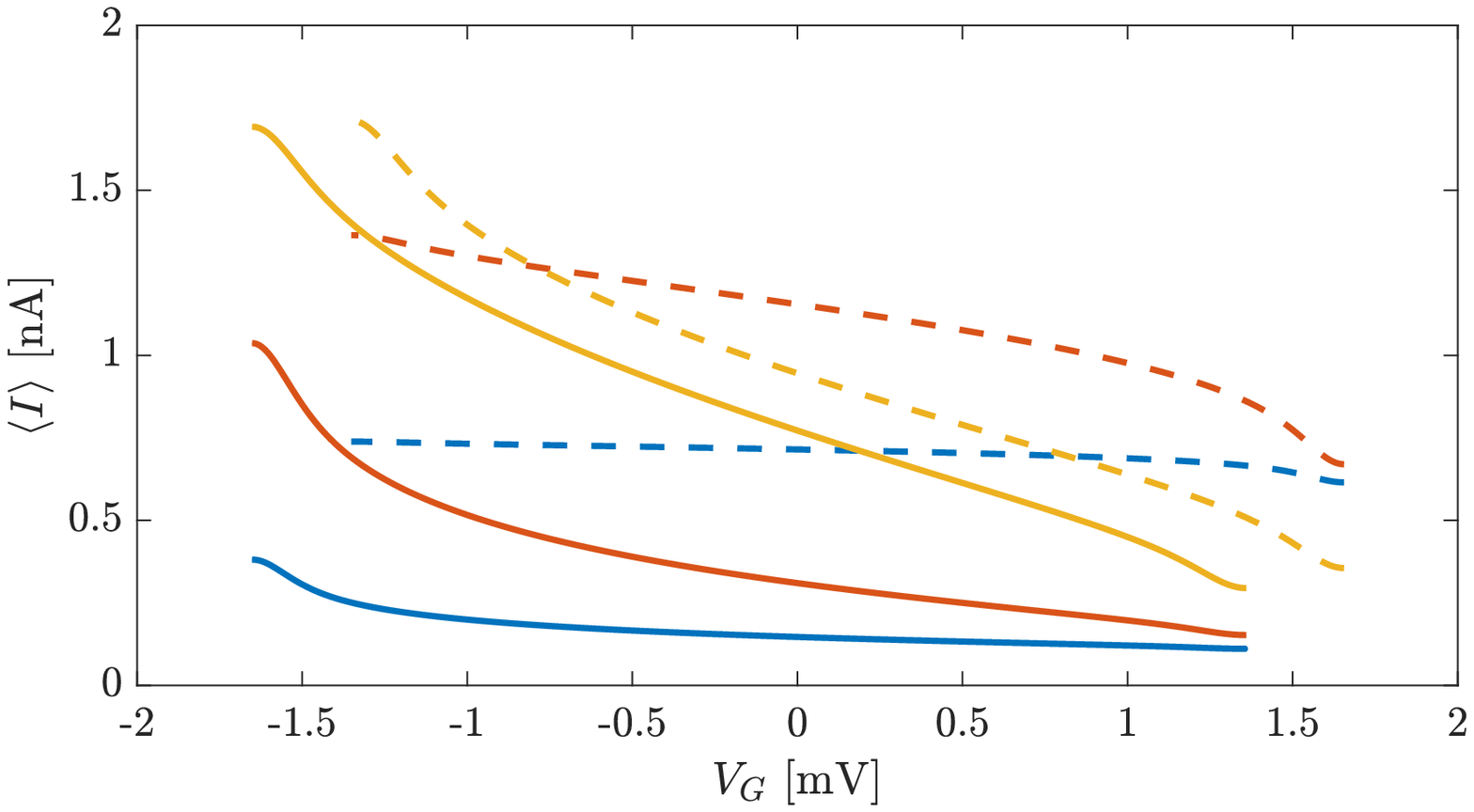}
    \put (12,40) {\large \textbf{(a)}}
    \end{overpic}
    \\
    \hfill\begin{overpic}[width = 0.99\linewidth, trim={0 0 0 0},clip]{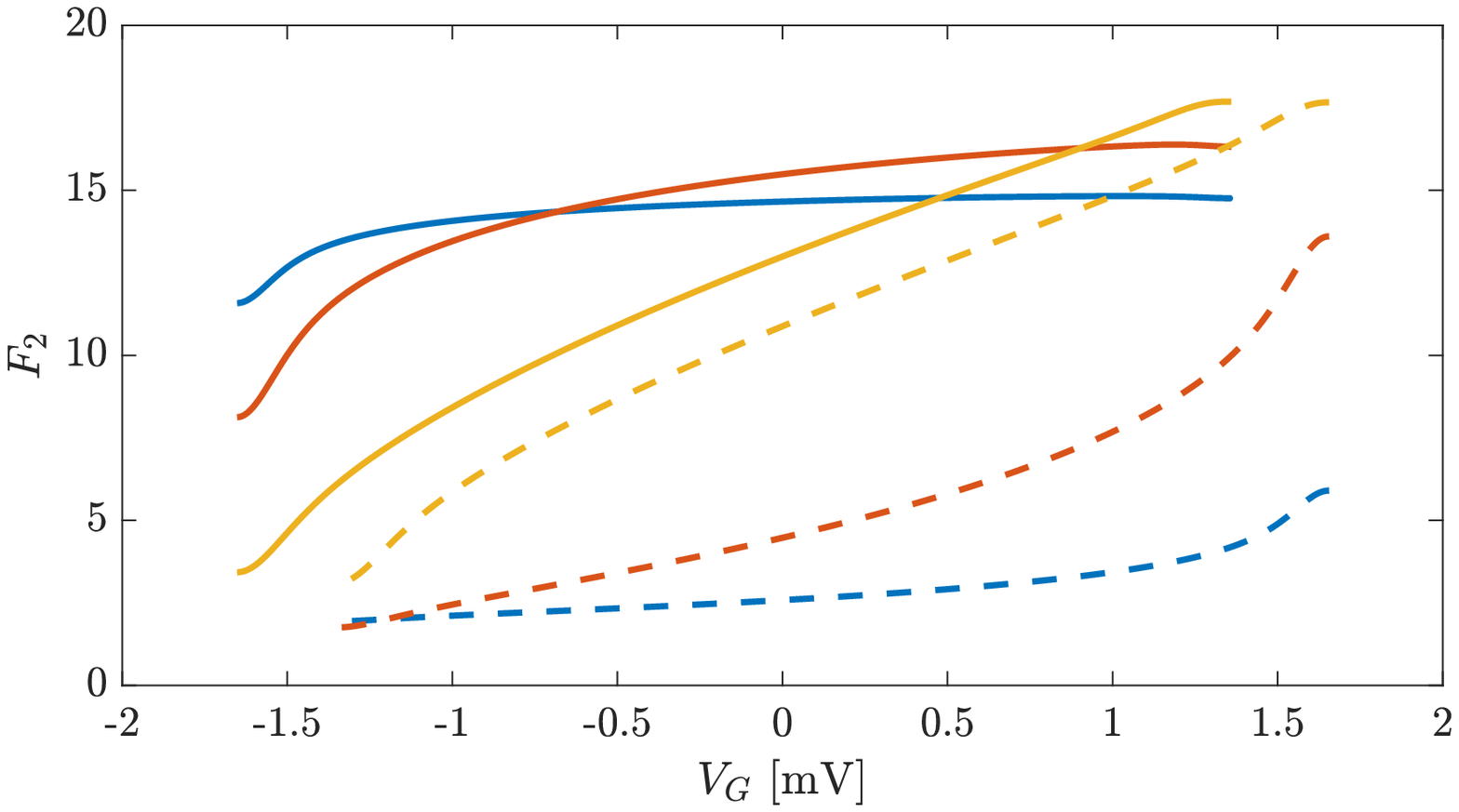}
    \put (11,50) {\large \textbf{(b)}}
    \end{overpic}
  \caption{ 
    The same as Fig.~\ref{fig:param_dp} but here we plot results for different values of $\Gamma_L =2,4,7\mu$eV (blue, red, yellow curves respectively) with the total rate $\Gamma_L +\Gamma_R = 14\mu$eV fixed at the experimentally observed value. Significant dependence of both quantities on the ratio $\Gamma_L/\Gamma_R$ is observed, but not as marked as the $\Delta \phi$ dependence.
  }
  \label{fig:param_dGLGR}
\end{figure}

In Fig.~\ref{fig:param_dGLGR} we investigate the current and noise characteristics as a functions of gate voltage with changes in the ratio of $\Gamma_L/\Gamma_R$ with total rate $\Gamma_L + \Gamma_R$ fixed.  We find that both quantities change significantly with $\Gamma_L/\Gamma_R$, and that the dependence of $F_2$ on $\Gamma_L/\Gamma_R$ is less marked than in its $\Delta \phi$ dependence.  The difference between forward and reverse properties with $\Gamma_L/\Gamma_R =1 $ stems from the inclusion of asymmetry factor $\eta \ne 1/2$, such that a small degree of rectification persists in this limit.

\begin{figure}[tb]
    \hfill \begin{overpic}[width = \linewidth, trim={0 0 0 0},clip]{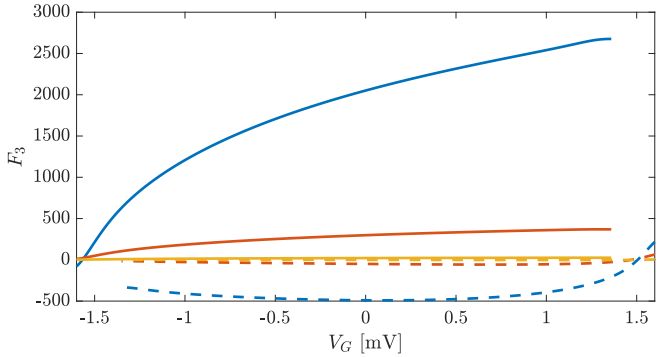}
    \end{overpic}
  \caption{ 
   Skewness Fano factor $F_3$ for the CNT-QD set-up as a function of gate voltage $V_G$. Same parameters and legend as Fig.~\ref{fig:param_dp}. With reverse bias, large positive values of $F_3$ are observed whereas at reverse bias, the skewness is negative across large portions on the gate-voltage range.
  }
  \label{fig:param_F3_dGLGR}
\end{figure}

Although the the shotnoise is the most readily accessible cumulant with current technology, we also plot in Fig.~\ref{fig:param_F3_dGLGR} the predicted skewness Fano factor $F_3$ as a function of gate voltage. As can be anticipated from the noise, the skewness Fano factor shows large values ($F_3 \approx 275 $ at $V_G=0$) for experimentally relevant parameters, and the results are very sensitive to small changes in the phase difference $\Delta \phi$.  
Also of interest is that, for the parameters found in Donarini \textit{et al.},  the forward-bias skewness is predicted to be negative throughout large parts of the gate-voltage range.

\section{Discussion}

In summary we have calculated the first three current cumulants of a CNT-QD containing a dark state. 
These cumulants are characterised by giant super-Poissonian Fano factors brought about electron bunching induced by the dark state.
The Fano factors show a strong  dependence on the phase difference $\Delta \phi$, and for $\Delta \phi \to 0$ with vanishing relaxation, the Fano factors diverge with $F_2>0$ and $F_3<0$.  This behaviour can be explained by noting that, in this limit, the system is  essentially bistable with a very long (in the limit, diverging) switching time between the states.
The phase-dependence, including the periodic divergence of of the Fano factors, is very similar to that reported Urban and K\"onig\cite{Urban09} (see also Li \textit{et al.} \cite{Li09}) for an Aharanov-Bohm interferometer with quantum dots in the arms.  In that context, the decisive phase is the flux through the interferometer, and as the flux varies, quantum-dot states couple and decouple from the leads, similar to the behaviour of the dark state here.
 
The bunching and bistability caused by the dark state are also responsible for giving the waiting time distributions the characteristic form found here.  These are composed of  an initial peak, corresponding to the conducting channel, followed by a extensive tail corresponding to long times long that the system spends trapped in the dark state. 

Using the parameters obtained for the CNT-QD measured in Ref.~\cite{Donarini19}, we predict that large Fano factors (of the order $F_2\approx 16$ and $F_3 \approx 275 $) should be observable in experiment, with a marked dependence on bias direction. We also predict that both signs of should be apparent in the measurement of the skewness Fano factor.  
The predicted sensitivity of the noise Fano Factor means that its measurement will be able to to more robustly determine the critical parameter $\Delta \phi$.  Finally, we point out that measurements of noise, and higher-order statistics, provide a more detailed test of the mechanisms at play in electron transport \cite{Kiesslich2007,Kurzmann2019}, and comparison with the predictions made here might reveal need for modifications to this dark-state model of CNT-QD transport.

\begin{acknowledgments}
The authors would like to thank Andrea Donarini and Milena Grifoni for helpful discussions.
\end{acknowledgments}

\appendix

\section{CNT-QD LIOUVILLIAN \label{AP:W}}
For the CNT-QD model, the density matrix has five relevant entries, which we organise into the vector $\kett{\rho} = \left(\rho_{00},\rho_{ll},\rho_{-l-l},\rho_{l-l},\rho_{-ll}\right)^\mathrm{T}$, with  transpose $\mathrm{T}$. In this basis, the Liouvillian for our problem reads
\begin{widetext}
\begin{equation}
        \mathcal{W}(\chi)=
    \begin{pmatrix}
        -4\Gamma_L & \Gamma_R e^{i\chi} & \Gamma_R e^{i\chi} & e^{-2i\phi_R+i \chi}\Gamma_{R} & e^{2i\phi_R+i \chi}\Gamma_{R} \\
        2\Gamma_L & -\Gamma_{R}-\frac{\Gamma_\mathrm{Rel}}{2} & \frac{\Gamma_\mathrm{Rel}}{2} & -(\frac{\Gamma_R}{2}e^{-2i\phi_R}-i\Tilde{\omega}^*) & -(\frac{\Gamma_R}{2}e^{2i\phi_R}+i\Tilde{\omega}) \\
        2\Gamma_L & \frac{\Gamma_\mathrm{Rel}}{2} & -\Gamma_{R}-\frac{\Gamma_\mathrm{Rel}}{2} & -(\frac{\Gamma_R}{2}e^{-2i\phi_R}+i\Tilde{\omega}^*) & -(\frac{\Gamma_R}{2}e^{2i\phi_R}-i\Tilde{\omega}) \\
        2\Gamma_{L}e^{2i\phi_L} & -(\frac{\Gamma_R}{2}e^{2i\phi_R}-i\Tilde{\omega}) & -(\frac{\Gamma_R}{2}e^{2i\phi_R}+i\Tilde{\omega}) & -(\Gamma_{R}+\Gamma_\mathrm{Rel}) & 0 \\
        2\Gamma_{L}e^{-2i\phi_L} & -(\frac{\Gamma_R}{2}e^{-2i\phi_R}+i\Tilde{\omega}^*) & -(\frac{\Gamma_R}{2}e^{-2i\phi_R}-i\Tilde{\omega}^*) & 0 &
        -(\Gamma_{R}+\Gamma_\mathrm{Rel})\\
    \end{pmatrix}
    \label{eq:CNTLiouvillian}
    ,
\end{equation}
where $\tilde{\omega}= \omega_L e^{2i\phi_L} + \omega_R e^{2i\phi_R}$.  Anticipating the next section, we have include here the counting-field factor $e^{i\chi}$ on all tunnel terms to the right lead.
\end{widetext}

\section{Counting Statistics Method \label{AP:FCS}}

Here we follow the full counting statistics formalism for Markovian master equations as described in e.g. Refs.~[\onlinecite{Bagrets,Flindt2004,Jauho2005,Flindt2008,Marcos}].  With the relevant density-matrix elements written into length-$N$ vector $\kett{\rho(t)}$, master equation~(\ref{EQ:QME}) can be written as  $\frac{d}{dt}\kett{\rho(t)} = \mathcal{W}\kett{\rho( t)}$ where $\mathcal{W}$ is the Liouville super-operator in matrix form.
Transport is then described by the $\chi$-resolved master equation
\begin{equation}
    \frac{d}{dt}\kett{\rho(\chi,t)} = \mathcal{W}(\chi)\kett{\rho(\chi, t)}.
    \label{eq:MME}
\end{equation}
where $\mathcal{W}(\chi)=\mathcal{W}_0 + e^{i\chi}\mathcal{J}$ is the Liouvillian decomposed into jump ($\mathcal{J}$) and non-jump ($\mathcal{W}_0$) parts, and where $\chi$ is the counting field. We have $\lim_{\chi \to 0}\mathcal{W}(\chi) = \mathcal{W} = \mathcal{W}_0 + \mathcal{J}$.  The stationary state of the system is given by
$
    \mathcal{W}\kett{\rho_\mathrm{stat}}=0
$,
which we assume to be unique. The left nullvector of $\mathcal{W}$, denoted $\braa{1}$, is normalised such that $\braakett{\mathds{1}}{\rho_{stat}}=1$.  Multiplication with this vector corresponds to taking the trace of the density matrix.  The expectation-value of a general Liouville-space operator $\mathcal{A}$ acting on state $\rho$ is thus is given by $\expec{\expec{\mathcal{A}}} = \braa{\mathds{1}}\mathcal{A} \kett{\rho_{stat}}$.  We also define $\mathcal{P} = \kett{\rho_{stat}}\braa{\mathds{1}}$ as the stationary-state projection matrix, and $\mathcal{R}$ the pseudo-inverse of $\mathcal{W}$.

The generating function of the cumulants of transferred charge is given by  $\mathcal{F}(\chi,t) = \lambda_0(\chi) t$, where $\lambda_0$ is that eigenvector of $ \mathcal{W}(\chi)$ which reverts to zero in the $\chi \to 0$ limit.  The current cumulants are then given by 
\begin{equation}
  \expec{I^k}_c =\frac{1}{t} \left. \frac{\partial^k}{\partial (i\chi)^k}\mathcal{F}(\chi,t)  \right|_0
  .
\end{equation}
Practically, however, finding the eigenvalues of $\mathcal{W}(\chi)$ is challenging for all but the simplest models, and an alternative approach is to expand the generating function to obtain explicit expressions for a finite set of cumulants.  The expressions for the first three current cumulants read ($e=1$):
\begin{eqnarray}
    \expec{I}_c &=& \expec{\expec{\mathcal{J}}};
    \nonumber\\
    \expec{I^2}_c &=& \expec{\expec{\mathcal{J} +2\mathcal{J}\mathcal{R}\mathcal{J}}};
    \nonumber\\
    \expec{I^3}_c &=& \expec{\expec{\mathcal{J}+3\mathcal{J}\mathcal{R}\mathcal{J}
    +3\mathcal{J}\mathcal{R}\mathcal{J}
    \nonumber\\
    &&
    ~~~~
    +
    6\mathcal{J}\mathcal{R}[\mathcal{J}\mathcal{R}-\mathcal{R}\mathcal{J}\mathcal{P} ]\mathcal{J}}}.
\label{eq:CurrentCumulant}
\end{eqnarray}

\par
The waiting time distribution for a master equation can be expressed in this same language \cite{TBrandes}. In the case of the carbon nanotube quantum dot we are only concerned with consecutive jumps of one type (transfer to the right lead) and the system is of the ``single reset'' type. In this case, the waiting time distribution is given by
\begin{equation}
    \omega(\tau) = \frac{\expec{\expec{\mathcal{J} e^{\mathcal{W}_0\tau}\mathcal{J} }}}{\expec{\expec{\mathcal{J}}}}
    \label{eq:wtd}
    .
\end{equation}

\section{Single-level model at $\Delta\phi=0$ \label{AP:SRL}}
At $\Delta\phi=0$ and with $\Gamma_\mathrm{rel}=0$, transport through the system can be described by the $\chi$-resolved Liouvillian
\begin{equation}
    \mathcal{W} (\chi) = 
    \begin{pmatrix}
    -4 \Gamma_L & \Gamma_R e^{i\chi}\\
    4 \Gamma_L & -\Gamma_R \\
    \end{pmatrix}
    ,
\end{equation}
written in the basis of empty state $\ket{0}$ and coupled state $\ket{CS}$ populations. The resulting cumulant generating function is \cite{Gustav2,Gustav1}
\begin{equation}
    \mathcal{F} (\chi) = \frac{\Gamma t}{2} \bigg( -1 + \sqrt{a^2 + 4 \frac{\expec{I}}{\Gamma} e^{i\chi}} \bigg),
\end{equation}
in terms of the total rate $\Gamma = 4\Gamma_L + \Gamma_R$, asymmetry $a = (4\Gamma_L - \Gamma_R) / \Gamma$, and mean current
\begin{equation}
    \expec{I} = \frac{4 \Gamma_L \Gamma_R}{4 \Gamma_L + \Gamma_R}
    \label{EQ:SRLcurrent}.
\end{equation}
The second and third Fano factors of this model read
\begin{equation}
    F_2 = \frac{1+a^2}{2}
    ;\qquad
    F_3 = \frac{1+3a^4}{4}
    .
\end{equation}
Interestingly, whereas the current expression Eq.~(\ref{EQ:SRLcurrent}) is also valid for $\Gamma_\mathrm{Rel} \ne 0$, the higher cumulants differ significantly from the above.

\providecommand{\noopsort}[1]{}\providecommand{\singleletter}[1]{#1}%

\end{document}